\documentclass[twocolumn,showpacs,preprintnumbers]{revtex4}%
\usepackage{amssymb}
\usepackage{amsmath}
\usepackage{graphicx}
\usepackage{dcolumn}
\usepackage{bm}
\usepackage{xcolor}
\usepackage{ifthen}
\usepackage[normalem]{ulem}
\usepackage{amsfonts}%
\UseRawInputEncoding
\providecommand{\U}[1]{\protect\rule{.1in}{.1in}}
\providecommand{\U}[1]{\protect\rule{.1in}{.1in}}
\def\showal{1}
\newcommand{\al}[1]{\ifthenelse{\showal=1}{\textcolor{orange}{[[#1]]}}{}}

\newcommand{\eb}[1]{\ifthenelse{\showal=1}{\textcolor{cyan}{[[#1]]}}{}}
\begin{document}
\title{The mean field approximation and disentanglement}
\author{Eyal Buks}
\affiliation{Andrew and Erna Viterbi Department of Electrical Engineering, Technion, Haifa
32000, Israel}
\date{\today }

\begin{abstract}
The mean field approximation becomes applicable when entanglement is
sufficiently weak. We explore a nonlinear term that can be added to the
Schr\"{o}dinger equation without violating unitarity of the time evolution. We
find that the added term suppresses entanglement, without affecting the
evolution of any product state. The dynamics generated by the modified
Schr\"{o}dinger equation is explored for the case of a two-spin 1/2 system. We
find that for this example the added term strongly affects the dynamics when
the Hartmann Hahn matching condition is nearly satisfied.

\end{abstract}
\maketitle





\section{Introduction}

Consider a system made of two subsystems. Let $A=A^{\dag}$ ($B=B^{\dag}$) be a
measurable of the first (second) subsystem. In the mean field approximation
\cite{breuer2002theory,Drossel_217,Hicke_024401} it is assumed that
$\left\langle AB\right\rangle =\left\langle A\right\rangle \left\langle
B\right\rangle $ (angle brackets denote an expectation value). This
approximation is valid when entanglement is sufficiently small, and it becomes
exact for any product state. Let $Q\left(  \left\vert \psi\right\rangle
\right)  $ be the level of entanglement of a given ket state vector
$\left\vert \psi\right\rangle $. There are several different ways to quantify
entanglement \cite{Wootters_2245}, however, none is linear in $\left\vert
\psi\right\rangle $, i.e. generally $Q\left(  \left\vert \psi_{1}\right\rangle
+\left\vert \psi_{2}\right\rangle \right)  \neq Q\left(  \left\vert \psi
_{1}\right\rangle \right)  +Q\left(  \left\vert \psi_{2}\right\rangle \right)
$. On the other hand, the Schr\"{o}dinger equation is linear in $\left\vert
\psi\right\rangle $. Moreover, the Gorini Kossakowski Sudarshan Lindblad
equation (GKSL) master equation \cite{Fernengel_385701,Lindblad_119} is linear
in the density operator $\rho$. Hence, a process which gives rise to
disentanglement only, without affecting the dynamics of product states, cannot
be properly described with time evolution that is generated either by
Schr\"{o}dinger or GKSL equations.

Here we consider a modified Schr\"{o}dinger equation, which includes a
nonlinear term that suppresses entanglement. The proposed equation can be
constructed for any physical system whose Hilbert space has finite dimensionality.

Previously proposed nonlinear terms that can be added to the
Schr\"{o}dinger equation are reviewed in \cite{Rembielinski_012027}. Weinberg
has considered a class of nonlinear Schr\"{o}dinger equations, for which
combining subsystems is possible \cite{Weinberg_336,Weinberg_61}. The time
evolution of the probability density generated by a nonlinear Schr\"{o}dinger
equation has been studied using the Fokker-Planck equation in
\cite{Doebner_397}. Gauge inveriance in nonlinear Schr\"{o}dinger equations
has been explored in \cite{Doebner_3764}. In most previous proposals, the
purpose of the added nonlinear terms is to generate a spontaneous collapse
\cite{Tumulka_821,Gisin_2259,Grigorenko_1459,Bialynicki_62,Hossenfelder_139}.

On the hand, here we propose an added nonlinear term that generates
disentanglement, i.e. it suppresses entanglement without affecting the
dynamics of product states. We explore the effect of the added
disentanglement nonlinear term for the case of a two-spin 1/2 system. With
externally applied driving the two-spin 1/2 system can become unstable
\cite{Levi_053516,Buks_052217} when the Hartmann Hahn matching condition
\cite{Hartmann1962,Yang_1} is nearly satisfied. We find that in the same
region the added nonlinear disentangling term has a relatively large effect on
the dynamics.

\section{The Schmidt decomposition}

Consider a system composed of two subsystems labeled as '1' and '2',
respectively. The dimensionality of the Hilbert spaces of both subsystems,
which is denoted by $N_{1}$ and $N_{2}$, respectively, is assumed to be
finite. The system is in a normalized pure state vector $\left\vert
\psi\right\rangle $ given by%
\begin{equation}
\left\vert \psi\right\rangle =\mathcal{K}_{1}C\otimes\mathcal{K}%
_{2}^{\mathrm{T}}\;,
\end{equation}
where $C$ is a $N_{1}\times N_{2}$ matrix having entries $C_{k_{1},k_{2}}$,
matrix transposition is denoted by $\mathrm{T}$, the raw vectors
$\mathcal{K}_{1}$ and $\mathcal{K}_{1}$\ are given by%
\begin{align}
\mathcal{K}_{1}  &  =\left(  \left\vert k_{1}\right\rangle _{1},\left\vert
k_{2}\right\rangle _{1},\cdots,\left\vert k_{N_{1}}\right\rangle _{1}\right)
\;,\\
\mathcal{K}_{2}  &  =\left(  \left\vert k_{1}\right\rangle _{2},\left\vert
k_{2}\right\rangle _{2},\cdots,\left\vert k_{N_{2}}\right\rangle _{2}\right)
\;,
\end{align}
and $\left\{  \left\vert k_{1}\right\rangle _{1}\right\}  $ ($\left\{
\left\vert k_{2}\right\rangle _{2}\right\}  $) is an orthonormal basis
spanning the Hilbert space of subsystem '1' ('2').

The purity $P_{1}$ ($P_{2}$) is defined by $P_{1}=\operatorname{Tr}\rho
_{1}^{2}$ ($P_{2}=\operatorname{Tr}\rho_{2}^{2}$), where $\rho_{1}%
=\operatorname{Tr}_{2}\rho$\ ($\rho_{2}=\operatorname{Tr}_{1}\rho$) is the
reduced density operator of the first (second) subsystems. By employing the
Schmidt decomposition one finds that $P_{1}=P_{2}\equiv P$, and that the level
of entanglement $Q$, which is defined by $Q=1-P$, is given by [see Eq.
(8.107)) in \cite{Buks_QMLN}]%
\begin{equation}
Q=2%
{\displaystyle\sum\limits_{k_{1}^{\prime}<k_{1}^{\prime\prime}}}
{\displaystyle\sum\limits_{k_{2}^{\prime}<k_{2}^{\prime\prime}}}
\left\vert \left\langle \Psi_{k_{1}^{\prime},k_{1}^{\prime\prime}%
,k_{2}^{\prime},k_{2}^{\prime\prime}}\right.  \left\vert \psi\right\rangle
\right\vert ^{2}\;, \label{Q}%
\end{equation}
where the state $\left\langle \Psi_{k_{1}^{\prime},k_{1}^{\prime\prime}%
,k_{2}^{\prime},k_{2}^{\prime\prime}}\right\vert $, which depends on on the
matrix $C$ corresponding to a given state $\left\vert \psi\right\rangle $, is
given by (note that $\left\langle \Psi_{k_{1}^{\prime},k_{1}^{\prime\prime
},k_{2}^{\prime},k_{2}^{\prime\prime}}\right\vert $ is not normalized)%
\begin{equation}
\left\langle \Psi_{k_{1}^{\prime},k_{1}^{\prime\prime},k_{2}^{\prime}%
,k_{2}^{\prime\prime}}\right\vert =C_{_{k_{1}^{\prime\prime},k_{2}%
^{\prime\prime}}}\left\langle k_{1}^{\prime},k_{2}^{\prime}\right\vert
-C_{k_{1}^{\prime\prime},k_{2}^{\prime}}\left\langle k_{1}^{\prime}%
,k_{2}^{\prime\prime}\right\vert \;. \label{<Psi_k|}%
\end{equation}
Note that $Q=0$ for a product state, and that $Q$ is time independent when the
subsystems are decoupled (i.e. their mutual interaction vanishes).

As an example, consider a two spin 1/2 system (i.e. $N_{1}=N_{2}=2$) in a pure
state $\left\vert \psi\right\rangle $ given by%
\begin{equation}
\left\vert \psi\right\rangle =a\left\vert --\right\rangle +b\left\vert
-+\right\rangle +c\left\vert +-\right\rangle +d\left\vert ++\right\rangle \;.
\label{|psi> 2S}%
\end{equation}
For this case Eq. (\ref{Q}) yields $Q=2\left\vert \left\langle \Psi\right.
\left\vert \psi\right\rangle \right\vert ^{2}$, where [for this case the sum
in Eq. (\ref{Q}) contains a single term with $k_{1}^{\prime}=-$,
$k_{1}^{\prime\prime}=+$, $k_{2}^{\prime}=-$ and $k_{2}^{\prime\prime}=+$]%
\begin{equation}
\left\langle \Psi\right\vert =d\left\langle --\right\vert -c\left\langle
-+\right\vert \;, \label{Psi 2S}%
\end{equation}
hence $Q=2\left\vert ad-bc\right\vert ^{2}$. The following holds $Q\leq1/2$
provided that $\left\vert \psi\right\rangle $ is normalized
\cite{Wootters_2245}.

\section{Disentanglement}

As will be shown below, entanglement can be suppressed by adding appropriate
nonlinear terms to the Schr\"{o}dinger equation. Consider a modified
Schr\"{o}dinger equation for the ket vector $\left\vert \psi\right\rangle $
having the form%
\begin{equation}
\frac{\mathrm{d}}{\mathrm{d}t}\left\vert \psi\right\rangle =\left(
-i\hbar^{-1}\mathcal{H}+\gamma_{\mathrm{D}}M_{\mathrm{D}}\right)  \left\vert
\psi\right\rangle \;, \label{GSE}%
\end{equation}
where $\hbar$ is the Planck's constant, $\mathcal{H}$ is the Hamiltonian, the
rate $\gamma_{\mathrm{D}}$ is positive, the operator $M_{\mathrm{D}}$ is given
by%
\begin{equation}
M_{\mathrm{D}}=-\sqrt{\frac{\left\langle \Psi\right.  \left\vert
\Psi\right\rangle }{1-\left\langle \mathcal{P}\right\rangle }}\left(
\mathcal{P}-\left\langle \mathcal{P}\right\rangle \right)  \;, \label{Omega_D}%
\end{equation}
the projection operator $\mathcal{P}$ is given by%
\begin{equation}
\mathcal{P}=\frac{\left\vert \Psi\right\rangle \left\langle \Psi\right\vert
}{\left\langle \Psi\right.  \left\vert \Psi\right\rangle }\;,
\label{P projection}%
\end{equation}
the expectation value $\left\langle \mathcal{P}\right\rangle $ is given by%
\begin{equation}
\left\langle \mathcal{P}\right\rangle =\frac{\left\langle \psi\right\vert
\mathcal{P}\left\vert \psi\right\rangle }{\left\langle \psi\right.  \left\vert
\psi\right\rangle }=\frac{\left\vert \left\langle \Psi\right.  \left\vert
\psi\right\rangle \right\vert ^{2}}{\left\langle \Psi\right.  \left\vert
\Psi\right\rangle \left\langle \psi\right.  \left\vert \psi\right\rangle }\;,
\label{<P>}%
\end{equation}
where $\left\vert \Psi\right\rangle $ is a given ket vector.

Note that $M_{\mathrm{D}}\left\vert \psi\right\rangle =0$ provided that
$\left\langle \Psi\right.  \left\vert \psi\right\rangle =0$, thus the added
term has no effect when $\left\vert \psi\right\rangle $ is orthogonal to
$\left\vert \Psi\right\rangle $. On the other hand, any product state
$\left\vert \psi\right\rangle $ is orthogonal to all the vectors $\left\vert
\Psi_{k_{1}^{\prime},k_{1}^{\prime\prime},k_{2}^{\prime},k_{2}^{\prime\prime}%
}\right\rangle $ given by Eq. (\ref{<Psi_k|}). Thus, any added term having the
form $\gamma_{\mathrm{D}}M_{\mathrm{D}}$, which is constructed based on a
vector $\left\vert \Psi\right\rangle $ that is one of the vectors $\left\vert
\Psi_{k_{1}^{\prime},k_{1}^{\prime\prime},k_{2}^{\prime},k_{2}^{\prime\prime}%
}\right\rangle $ given by Eq. (\ref{<Psi_k|}), has no effect when $\left\vert
\psi\right\rangle $ represents a product state.

Using Eqs. (\ref{Omega_D}) and (\ref{P projection}) one finds that (note that
$\mathcal{P}^{2}=\mathcal{P}$)%
\begin{align}
\left\langle \psi\right\vert M_{\mathrm{D}}\left\vert \psi\right\rangle  &
=0\;,\label{OmD 1}\\
\left\langle \Psi\right\vert M_{\mathrm{D}}\left\vert \psi\right\rangle  &
=-\sqrt{\left\langle \Psi\right.  \left\vert \Psi\right\rangle \left(
1-\left\langle \mathcal{P}\right\rangle \right)  }\left\langle \Psi\right.
\left\vert \psi\right\rangle \;,\label{OmD 2}\\
\left\langle \psi\right\vert M_{\mathrm{D}}^{2}\left\vert \psi\right\rangle
&  =\left\vert \left\langle \Psi\right.  \left\vert \psi\right\rangle
\right\vert ^{2}\;. \label{OmD 3}%
\end{align}
The relation (\ref{OmD 1}) implies that $M_{\mathrm{D}}\left\vert
\psi\right\rangle $ is orthogonal to $\left\vert \psi\right\rangle $, and thus
the unitarity condition, which reads [see Eq. (\ref{GSE})] $0=\left(
\mathrm{d}/\mathrm{d}t\right)  \left\langle \psi\right.  \left\vert
\psi\right\rangle =i\hbar^{-1}\left\langle \psi\right\vert \mathcal{H}^{\dag
}-\mathcal{H}\left\vert \psi\right\rangle +2\gamma_{\mathrm{D}}%
\operatorname*{Re}\left\langle \psi\right\vert M_{\mathrm{D}}\left\vert
\psi\right\rangle $, is satisfied provided that the Hamiltonian is Hermitian,
i.e.$\mathcal{H}^{\dag}=\mathcal{H}$ (henceforth it is assumed that
$\mathcal{H}$ is Hermitian).

The modified Schr\"{o}dinger equation (\ref{GSE}) yields a modified Heisenberg
equation given by%
\begin{equation}
\frac{\mathrm{d}}{\mathrm{d}t}\left\langle \psi\right\vert O\left\vert
\psi\right\rangle =\frac{\left\langle \psi\right\vert \left[  O,\mathcal{H}%
\right]  \left\vert \psi\right\rangle }{i\hbar}+\gamma_{\mathrm{D}%
}\left\langle \psi\right\vert \left\{  O,M_{\mathrm{D}}\right\}  \left\vert
\psi\right\rangle \;, \label{GHE}%
\end{equation}
where $O=O^{\dag}$ is a given observable,that does not explicitly depend on
time, and where $\left\{  O,M_{\mathrm{D}}\right\}  =OM_{\mathrm{D}%
}+M_{\mathrm{D}}O$. For the case where $O=\mathcal{P}$ Eq. (\ref{GHE}) yields
[see Eq. (\ref{OmD 2})]%
\begin{align}
\frac{\mathrm{d}}{\mathrm{d}t}\frac{\left\vert \left\langle \Psi\right.
\left\vert \psi\right\rangle \right\vert ^{2}}{\left\langle \Psi\right.
\left\vert \Psi\right\rangle }  &  =\frac{\left\langle \psi\right\vert \left[
\mathcal{P},\mathcal{H}\right]  \left\vert \psi\right\rangle }{i\hbar
}\nonumber\\
&  -2\gamma_{\mathrm{D}}\sqrt{\left\langle \Psi\right.  \left\vert
\Psi\right\rangle \left(  1-\left\langle \mathcal{P}\right\rangle \right)
}\frac{\left\vert \left\langle \Psi\right.  \left\vert \psi\right\rangle
\right\vert ^{2}}{\left\langle \Psi\right.  \left\vert \Psi\right\rangle
}\;.\nonumber\\
&  \label{d/dt Psi psi=}%
\end{align}
An upper bound can be derived for the first term on the right hand side of Eq.
(\ref{d/dt Psi psi=}) using the uncertainty principle $\left\vert \left\langle
\left[  \mathcal{P},\mathcal{H}\right]  \right\rangle \right\vert ^{2}%
\leq4\left\langle \left(  \mathcal{P}-\left\langle \mathcal{P}\right\rangle
\right)  ^{2}\right\rangle \left\langle \left(  \mathcal{H}-\left\langle
\mathcal{H}\right\rangle \right)  ^{2}\right\rangle \leq\left\langle \left(
\mathcal{H}-\left\langle \mathcal{H}\right\rangle \right)  ^{2}\right\rangle $
[note that $\left\langle \left(  \mathcal{P}-\left\langle \mathcal{P}%
\right\rangle \right)  ^{2}\right\rangle =\left\langle \mathcal{P}%
^{2}\right\rangle -\left\langle \mathcal{P}\right\rangle ^{2}$, $\mathcal{P}%
^{2}=\mathcal{P}$, $0\leq\left\langle \mathcal{P}\right\rangle \leq1$ and
$0\leq4x\left(  1-x\right)  \leq1$ for $0\leq x\leq1$]. The second term on the
right hand side of Eq. (\ref{d/dt Psi psi=}) represents a rotation of the ket
vector $\left\vert \psi\right\rangle $ away from the ket vector $\left\vert
\Psi\right\rangle $. This rotation gives rise to disentanglement when
$\left\vert \Psi\right\rangle $ is chosen to be one of the vectors $\left\vert
\Psi_{k_{1}^{\prime},k_{1}^{\prime\prime},k_{2}^{\prime},k_{2}^{\prime\prime}%
}\right\rangle $ given by Eq. (\ref{<Psi_k|}).

\section{Two-spin system}

As an example, consider two two-level systems (TLS) having a mutual coupling
that is characterized by a coupling coefficient $g$. The first TLS, which is
labelled as '$\mathrm{a}$', has a relatively low angular frequency
$\omega_{\mathrm{a}}$ in comparison with the angular frequency $\omega
_{\mathrm{b}}$ of the second TLS, which is labelled as '$\mathrm{b}$', and
which is externally driven. The Hamiltonian $\mathcal{H}$ of the closed system
is given by%
\begin{align}
\mathcal{H}  &  =\omega_{\mathrm{a}}S_{\mathrm{az}}+\omega_{\mathrm{b}%
}S_{\mathrm{bz}}+\frac{\omega_{1}\left(  S_{\mathrm{b+}}+S_{\mathrm{b-}%
}\right)  }{2}\nonumber\\
&  +g\hbar^{-1}\left(  S_{\mathrm{a+}}+S_{\mathrm{a-}}\right)  S_{\mathrm{bz}%
}\;,\nonumber\\
&
\end{align}
where the driving amplitude and angular frequency are denoted by $\omega_{1}$
and $\omega_{\mathrm{p}}=\omega_{\mathrm{b}}+\Delta$, respectively ($\Delta$
is the driving detuning), the operators $S_{\mathrm{a\pm}}$ are given by
$S_{\mathrm{a\pm}}=S_{\mathrm{ax}}\pm iS_{\mathrm{ay}}$, and the rotated
operators $S_{\mathrm{b\pm}}$ are given by $S_{\mathrm{b\pm}}=\left(
S_{\mathrm{bx}}\pm iS_{\mathrm{by}}\right)  e^{\pm i\omega_{\mathrm{p}}t}$. In
the basis $\left\{  \left\vert --\right\rangle ,\left\vert -+\right\rangle
,\left\vert +-\right\rangle ,\left\vert ++\right\rangle \right\}  $ the matrix
representation of the Hamiltonian is given by $\hbar\Omega$, where the matrix
$\Omega$ is given by%
\begin{equation}
\Omega=\left(
\begin{array}
[c]{cccc}%
\frac{-\omega_{\mathrm{a}}-\omega_{\mathrm{b}}}{2} & \frac{\omega_{1}}%
{2}e^{i\omega_{\mathrm{p}}t} & -\frac{g}{2} & 0\\
\frac{\omega_{1}}{2}e^{-i\omega_{\mathrm{p}}t} & \frac{-\omega_{\mathrm{a}%
}+\omega_{\mathrm{b}}}{2} & 0 & \frac{g}{2}\\
-\frac{g}{2} & 0 & \frac{\omega_{\mathrm{a}}-\omega_{\mathrm{b}}}{2} &
\frac{\omega_{1}}{2}e^{i\omega_{\mathrm{p}}t}\\
0 & \frac{g}{2} & \frac{\omega_{1}}{2}e^{-i\omega_{\mathrm{p}}t} &
\frac{\omega_{\mathrm{a}}+\omega_{\mathrm{b}}}{2}%
\end{array}
\right)  \;. \label{Omega}%
\end{equation}
The unitary transformations $U_{1}=1_{\mathrm{a}}\otimes u_{\mathrm{b1}}$ and
$U_{2}=1_{\mathrm{a}}\otimes u_{\mathrm{b2}}$ are successively employed below,
where $1_{\mathrm{a}}$ is the identity operator of TLS a. For the first one,
which transforms TLS b to a frame rotated at the angular driving frequency
$\omega_{\mathrm{p}}$, the matrix representation of $u_{\mathrm{b1}}$ is given
by%
\begin{equation}
u_{\mathrm{b1}}\dot{=}\left(
\begin{array}
[c]{cc}%
e^{\frac{i\omega_{\mathrm{p}}t}{2}} & 0\\
0 & e^{-\frac{i\omega_{\mathrm{p}}t}{2}}%
\end{array}
\right)  \;,
\end{equation}
and the corresponding transformed matrix $\Omega^{\prime}$ is given by [see
Eq. (\ref{Omega}) and Eq. (6.329) in \cite{Buks_QMLN}]%
\begin{equation}
\Omega^{\prime}=\left(
\begin{array}
[c]{cccc}%
\frac{-\omega_{\mathrm{a}}+\Delta}{2} & \frac{\omega_{1}}{2} & -\frac{g}{2} &
0\\
\frac{\omega_{1}}{2} & \frac{-\omega_{\mathrm{a}}-\Delta}{2} & 0 & \frac{g}%
{2}\\
-\frac{g}{2} & 0 & \frac{\omega_{\mathrm{a}}+\Delta}{2} & \frac{\omega_{1}}%
{2}\\
0 & \frac{g}{2} & \frac{\omega_{1}}{2} & \frac{\omega_{\mathrm{a}}-\Delta}{2}%
\end{array}
\right)  \;. \label{Omega'}%
\end{equation}
For the second transformation the matrix representation of $u_{\mathrm{b2}}$
is given by%
\begin{equation}
u_{\mathrm{b2}}\dot{=}\left(
\begin{array}
[c]{cc}%
\cos\frac{\theta}{2} & \sin\frac{\theta}{2}\\
-\sin\frac{\theta}{2} & \cos\frac{\theta}{2}%
\end{array}
\right)  \;,
\end{equation}
where%
\begin{equation}
\tan\theta=-\frac{\omega_{1}}{\Delta}\;,
\end{equation}
and the corresponding transformed matrix $\Omega^{\prime\prime}$ is given by
[see Eq. (\ref{Omega'})]%
\begin{equation}
\Omega^{\prime\prime}=\left(
\begin{array}
[c]{cccc}%
\frac{-\omega_{\mathrm{a}}+\omega_{\mathrm{R}}}{2} & 0 & -\frac{g\Delta
}{2\omega_{\mathrm{R}}} & \frac{g\omega_{1}}{2\omega_{\mathrm{R}}}\\
0 & \frac{-\omega_{\mathrm{a}}-\omega_{\mathrm{R}}}{2} & \frac{g\omega_{1}%
}{2\omega_{\mathrm{R}}} & \frac{g\Delta}{2\omega_{\mathrm{R}}}\\
-\frac{g\Delta}{2\omega_{\mathrm{R}}} & \frac{g\omega_{1}}{2\omega
_{\mathrm{R}}} & \frac{\omega_{\mathrm{a}}+\omega_{\mathrm{R}}}{2} & 0\\
\frac{g\omega_{1}}{2\omega_{\mathrm{R}}} & \frac{g\Delta}{2\omega_{\mathrm{R}%
}} & 0 & \frac{\omega_{\mathrm{a}}-\omega_{\mathrm{R}}}{2}%
\end{array}
\right)  \;, \label{Omega''}%
\end{equation}
where $\omega_{\mathrm{R}}$, which is given by%
\begin{equation}
\omega_{\mathrm{R}}=\sqrt{\omega_{1}^{2}+\Delta^{2}}\;,
\end{equation}
is the Rabi angular frequency.

The matrix $\Omega^{\prime\prime}$ in the limit where the TLSs are decoupled
is denoted by $\Omega_{0}^{\prime\prime}$, i.e. $\Omega_{0}^{\prime\prime
}=\lim_{g\rightarrow0}\Omega^{\prime\prime}$. The first and forth energy
eigenvalues of $\Omega_{0}^{\prime\prime}$ become degenerate when the Hartmann
Hahn matching condition $\omega_{\mathrm{a}}=\omega_{\mathrm{R}}$ is
satisfied. Consequently, the effect of the coupling becomes relatively strong
when $\omega_{\mathrm{a}}\simeq\omega_{\mathrm{R}}$. In this region the
problem can be simplified by employing a truncation approximation into the
subspace spanned by the transformed states $\left\vert --\right\rangle $ and
$\left\vert ++\right\rangle $ [first and forth vectors of the basis that is
used for constructing the matrix $\Omega^{\prime\prime}$ given by Eq.
(\ref{Omega''})]. In this approximation the $4\times4$ matrix $\Omega
^{\prime\prime}$ is replaced by the $2\times2$ truncated matrix $\Omega
_{\mathrm{T}}^{\prime\prime}$, which is given by%
\begin{equation}
\Omega_{\mathrm{T}}^{\prime\prime}=\left(
\begin{array}
[c]{cc}%
\frac{\omega_{\mathrm{R}}-\omega_{\mathrm{a}}}{2} & \frac{g\omega_{1}}%
{2\omega_{\mathrm{R}}}\\
\frac{g\omega_{1}}{2\omega_{\mathrm{R}}} & -\frac{\omega_{\mathrm{R}}%
-\omega_{\mathrm{a}}}{2}%
\end{array}
\right)  =\boldsymbol{\omega}\cdot\boldsymbol{\sigma}\;, \label{Omega_T''}%
\end{equation}
where the effective magnetic field vector $\boldsymbol{\omega}$ is given by%
\begin{equation}
\boldsymbol{\omega}=\left(  \frac{g\omega_{1}}{2\omega_{\mathrm{R}}}%
,0,\frac{\omega_{\mathrm{R}}-\omega_{\mathrm{a}}}{2}\right)  \equiv\left(
\omega_{x},0,\omega_{z}\right)  \;, \label{k vector}%
\end{equation}
and the components of the Pauli matrix vector $\boldsymbol{\sigma}=\left(
\sigma_{x},\sigma_{y},\sigma_{z}\right)  $\ are given by%
\begin{equation}
\sigma_{x}=\left(
\begin{array}
[c]{cc}%
0 & 1\\
1 & 0
\end{array}
\right)  ,\;\sigma_{y}=\left(
\begin{array}
[c]{cc}%
0 & -i\\
i & 0
\end{array}
\right)  ,\;\sigma_{z}=\left(
\begin{array}
[c]{cc}%
1 & 0\\
0 & -1
\end{array}
\right)  \;. \label{Pauli matrix vector}%
\end{equation}
The notation given by Eq. (\ref{|psi> 2S}) is employed below for the
transformed state $\left\vert \psi^{\prime\prime}\right\rangle $. It is
henceforth assumed that $\left\vert \psi^{\prime\prime}\right\rangle $ is
normalized, i.e. $\left\vert a\right\vert ^{2}+\left\vert d\right\vert ^{2}=1$
(note that $b=c=0$ for the truncation approximation). The polarization vector
$\mathbf{P}$ is given by $\mathbf{P}=\left(  \left\langle \psi\right\vert
\sigma_{x}\left\vert \psi\right\rangle ,\left\langle \psi\right\vert
\sigma_{y}\left\vert \psi\right\rangle ,\left\langle \psi\right\vert
\sigma_{z}\left\vert \psi\right\rangle \right)  $, where $P_{x}=d^{\ast
}a+a^{\ast}d$, $P_{y}=i\left(  d^{\ast}a-a^{\ast}d\right)  $ and
$P_{z}=\left\vert a\right\vert ^{2}-\left\vert d\right\vert ^{2}$. Note that
$\mathbf{P}\cdot\mathbf{P}=1$ provided that $\left\vert \psi^{\prime\prime
}\right\rangle $ is normalized.

In the truncation approximation $M_{\mathrm{D}}$ is replaced by the $2\times2$
matrix $M_{\mathrm{DT}}$, which is given by [see Eqs. (\ref{Psi 2S}) and
(\ref{Omega_D})]%
\begin{equation}
M_{\mathrm{DT}}=\left(
\begin{array}
[c]{cc}%
-\left\vert d\right\vert ^{2} & 0\\
0 & \left\vert a\right\vert ^{2}%
\end{array}
\right)  =\frac{\left\vert a\right\vert ^{2}-\left\vert d\right\vert ^{2}}%
{2}-\frac{\sigma_{z}}{2}\;. \label{M_DT}%
\end{equation}
The following holds $\left\langle \psi^{\prime\prime}\right\vert
M_{\mathrm{DT}}^{2}\left\vert \psi^{\prime\prime}\right\rangle =\left\vert
ad\right\vert ^{2}=Q/2$, where $Q$ is the level of entanglement [recall that
$\left\vert a\right\vert ^{2}+\left\vert d\right\vert ^{2}=1$, and compare
with Eq. (\ref{OmD 3})].

With the help of the identity%
\begin{equation}
\left(  \boldsymbol{\sigma}\cdot\mathbf{a}\right)  \left(
\boldsymbol{\sigma}\cdot\mathbf{b}\right)  =\mathbf{a}\cdot\mathbf{b}%
+i\boldsymbol{\sigma}\cdot\left(  \mathbf{a}\times\mathbf{b}\right)  \;,
\label{(sigma dot a)(sigma dot b)}%
\end{equation}
where $\mathbf{a}$ and $\mathbf{b}$ are given vectors, one finds that [recall
the vector identity $\mathbf{A}\cdot\left(  \mathbf{B}\times\mathbf{C}\right)
=\mathbf{B}\cdot\left(  \mathbf{C}\times\mathbf{A}\right)  =\mathbf{C}%
\cdot\left(  \mathbf{A}\times\mathbf{B}\right)  $, and see Eqs. (\ref{GHE}),
(\ref{Omega_T''}) and (\ref{M_DT})]%
\begin{equation}
\frac{\mathrm{d}\mathbf{P}}{\mathrm{d}t}=2\boldsymbol{\omega}\times
\mathbf{P}+\gamma_{\mathrm{D}}\mathbf{V}_{\mathrm{D}}\;, \label{eom P V1}%
\end{equation}
where the vector $\mathbf{V}_{\mathrm{D}}$ is given by [note that $\left\{
\sigma_{x},\sigma_{z}\right\}  =\left\{  \sigma_{y},\sigma_{z}\right\}  =0$
and $\left\{  \sigma_{z},\sigma_{z}\right\}  =2$, see Eq.
(\ref{(sigma dot a)(sigma dot b)})]%
\begin{equation}
\mathbf{V}_{\mathrm{D}}=P_{z}\mathbf{P}-\mathbf{\hat{z}}\;,
\end{equation}
where $\mathbf{\hat{z}}$ is a unit vector in the $z$ direction. The following
holds $\mathbf{P}\cdot\mathbf{V}_{\mathrm{D}}=0$ [compare with Eq.
(\ref{OmD 1})] and $\mathbf{V}_{\mathrm{D}}\cdot\mathbf{V}_{\mathrm{D}%
}=1-P_{z}^{2}=4\left\vert ad\right\vert ^{2}=2Q$\ [recall that $\mathbf{P}%
\cdot\mathbf{P}=1$ and $P_{z}=\left\vert a\right\vert ^{2}-\left\vert
d\right\vert ^{2}$, and compare with Eq. (\ref{OmD 3})]. For a normalized
$\mathbf{P}$ the following holds $\mathbf{V}_{\mathrm{D}}=\left(
\mathbf{\hat{z}}\times\mathbf{P}\right)  \times\mathbf{P}$, hence Eq.
(\ref{eom P V1}) can be rewritten as%
\begin{equation}
\frac{\mathrm{d}\mathbf{P}}{\mathrm{d}t}=\left(  2\boldsymbol{\omega}+\gamma
_{\mathrm{D}}\mathbf{\hat{z}}\times\mathbf{P}\right)  \times\mathbf{P}\;.
\label{eom P}%
\end{equation}
Note that when \boldsymbol{$\omega$}$\mathbf{\parallel\hat{z}}$ the equation
of motion (\ref{eom P}) is similar to the Landau--Lifshitz equation for the
time evolution of the magnetization vector of a ferromagnet. However, the
condition \boldsymbol{$\omega$}$\mathbf{\parallel\hat{z}}$, which is satisfied
only when the driving amplitude $\omega_{1}$ vanishes [see Eq. (\ref{k vector}%
)], is inconsistent with the assumption that the Hartmann Hahn matching
condition is nearly satisfied.

Let $\mathbf{P}_{0}=\left(  \sin\theta\cos\phi,\sin\theta\sin\phi,\cos
\theta\right)  $ be a fixed point of Eq. (\ref{eom P}) (i.e. $\mathrm{d}%
\mathbf{P}/\mathrm{d}t=0$ at $\mathbf{P}_{0}$). Using Eq. (\ref{eom P}) one
finds that the angle $\theta$ for the fixed point $\mathbf{P}_{0}$ can be
found by solving%
\begin{equation}
\frac{\gamma_{\mathrm{D}}^{2}}{4\left(  \omega_{x}^{2}+\omega_{z}^{2}\right)
}=\frac{\sin\left(  \theta_{\mathrm{H}}-\theta\right)  \sin\left(
\theta_{\mathrm{H}}+\theta\right)  }{\sin^{2}\theta\cos^{2}\theta}\;,
\label{theta P0}%
\end{equation}
where%
\begin{equation}
\tan\theta_{\mathrm{H}}=\frac{\omega_{x}}{\omega_{z}}\;.
\end{equation}
According to Eq. (\ref{theta P0}), in the absence of disentanglement, i.e.
when $\gamma_{\mathrm{D}}=0$, the vector $\mathbf{P}_{0}$ is parallel to the
vector \boldsymbol{$\omega$} (\ref{k vector}), whereas in the opposite limit
of strong disentanglement, i.e. when $\gamma_{\mathrm{D}}^{2}\gg\omega_{x}%
^{2}+\omega_{z}^{2}$, the vector $\mathbf{P}_{0}$ becomes nearly parallel to
$\mathbf{\hat{z}}$ (i.e. the state represented by the fixed point
$\mathbf{P}_{0}$ nearly becomes a product state).

\section{Summary}

In summary, the modified Schr\"{o}dinger equation suppresses entanglement
without violating unitarity. Future study will be devoted to the rich
nonlinear dynamics that is generated by the added disentanglement term. Some
outstanding questions, which were left outside the scope of the current
manuscript, are briefly mentioned below. How a given system should be divided
into two (or perhaps more) subsystems? In traditional quantum mechanics such a
division is generally not unique. How to determine the values of the
disentanglement rates? The hypothesis that these rates remain finite even when
the subsystems become decoupled is likely to be inconsistent with the
principle of causality \cite{Bassi_055027,Jordan_022101} (recall that in
traditional quantum mechanics the level of entanglement $Q$ becomes time
independent when the subsystems are decoupled). Recently, it was
shown that when a condition, called 'convex quasilinearity' is satisfied by a
given nonlinear Schr\"{o}dinger equation, violation of the causality principle
becomes impossible \cite{Rembielinski_012027}. Upper bound
imposed upon the disentanglement rates can be derived from experimental
observations of quantum entanglement.

The existence of quantum entanglement has been conclusively demonstrated in
many different physical systems. On the other hand, entanglement can be held
responsible for a fundamental self-inconsistency related to the quantum to
classical transition \cite{Penrose_4864,Leggett_939,Leggett_022001}, which was
first introduced by Schr\"{o}dinger \cite{Schrodinger_807} (this
self-inconsistency is commonly known as the problem of quantum measurement).
Exploring possible mechanisms of disentanglement may help resolving this
long-standing self-inconsistency.

\section{Acknowledgments}

We thank Ido Kaminer and Eliahu Cohen for useful discussions. This work is
supported by the Israeli Science foundation.

\bibliographystyle{ieeepes}
\bibliography{acompat,Eyal_Bib}

\newif\ifabfull\abfulltrue
\begin{thebibliography}{10}

\bibitem{breuer2002theory}
Heinz-Peter Breuer, Francesco Petruccione, et~al.,
\newblock {\em The theory of open quantum systems},
\newblock Oxford University Press on Demand, 2002.

\bibitem{Drossel_217}
Barbara Drossel,
\newblock ``What condensed matter physics and statistical physics teach us
  about the limits of unitary time evolution'',
\newblock {\em Quantum Studies: Mathematics and Foundations}, vol. 7, no. 2,
  pp. 217--231, 2020.

\bibitem{Hicke_024401}
C~Hicke and MI~Dykman,
\newblock ``Classical dynamics of resonantly modulated large-spin systems'',
\newblock {\em Physical Review B}, vol. 78, no. 2, pp. 024401, 2008.

\bibitem{Wootters_2245}
William~K Wootters,
\newblock ``Entanglement of formation of an arbitrary state of two qubits'',
\newblock {\em Physical Review Letters}, vol. 80, no. 10, pp. 2245, 1998.

\bibitem{Fernengel_385701}
Bernd Fernengel and Barbara Drossel,
\newblock ``Bifurcations and chaos in nonlinear lindblad equations'',
\newblock {\em Journal of Physics A: Mathematical and Theoretical}, vol. 53,
  no. 38, pp. 385701, 2020.

\bibitem{Lindblad_119}
Goran Lindblad,
\newblock ``On the generators of quantum dynamical semigroups'',
\newblock {\em Communications in Mathematical Physics}, vol. 48, no. 2, pp.
  119--130, 1976.

\bibitem{Rembielinski_012027}
Jakub Rembieli{\'n}ski and Pawe{\l} Caban,
\newblock ``Nonlinear evolution and signaling'',
\newblock {\em Physical Review Research}, vol. 2, no. 1, pp. 012027, 2020.

\bibitem{Weinberg_336}
Steven Weinberg,
\newblock ``Testing quantum mechanics'',
\newblock {\em Annals of Physics}, vol. 194, no. 2, pp. 336--386, 1989.

\bibitem{Weinberg_61}
Steven Weinberg,
\newblock ``Precision tests of quantum mechanics'',
\newblock in {\em THE OSKAR KLEIN MEMORIAL LECTURES 1988--1999}, pp. 61--68.
  World Scientific, 2014.

\bibitem{Doebner_397}
H-D Doebner and Gerald~A Goldin,
\newblock ``On a general nonlinear schr{\"o}dinger equation admitting diffusion
  currents'',
\newblock {\em Physics Letters A}, vol. 162, no. 5, pp. 397--401, 1992.

\bibitem{Doebner_3764}
H-D Doebner and Gerald~A Goldin,
\newblock ``Introducing nonlinear gauge transformations in a family of
  nonlinear schr{\"o}dinger equations'',
\newblock {\em Physical Review A}, vol. 54, no. 5, pp. 3764, 1996.

\bibitem{Tumulka_821}
Roderich Tumulka,
\newblock ``A relativistic version of the ghirardi--rimini--weber model'',
\newblock {\em Journal of Statistical Physics}, vol. 125, no. 4, pp. 821--840,
  2006.

\bibitem{Gisin_2259}
Nicolas Gisin,
\newblock ``A simple nonlinear dissipative quantum evolution equation'',
\newblock {\em Journal of Physics A: Mathematical and General}, vol. 14, no. 9,
  pp. 2259, 1981.

\bibitem{Grigorenko_1459}
AN~Grigorenko,
\newblock ``Measurement description by means of a nonlinear schrodinger
  equation'',
\newblock {\em Journal of Physics A: Mathematical and General}, vol. 28, no. 5,
  pp. 1459, 1995.

\bibitem{Bialynicki_62}
Iwo Bialynicki-Birula and Jerzy Mycielski,
\newblock ``Nonlinear wave mechanics'',
\newblock {\em Annals of Physics}, vol. 100, no. 1-2, pp. 62--93, 1976.

\bibitem{Hossenfelder_139}
Sabine Hossenfelder and Tim Palmer,
\newblock ``Rethinking superdeterminism'',
\newblock {\em Frontiers in Physics}, vol. 8, pp. 139, 2020.

\bibitem{Levi_053516}
Roei Levi, Sergei Masis, and Eyal Buks,
\newblock ``Instability in the hartmann-hahn double resonance'',
\newblock {\em Phys. Rev. A}, vol. 102, pp. 053516, Nov 2020.

\bibitem{Buks_052217}
Eyal Buks and Dvir Schwartz,
\newblock ``Stability of the grabert master equation'',
\newblock {\em Physical Review A}, vol. 103, no. 5, pp. 052217, 2021.

\bibitem{Hartmann1962}
SR~Hartmann and EL~Hahn,
\newblock ``Nuclear double resonance in the rotating frame'',
\newblock {\em Physical Review}, vol. 128, no. 5, pp. 2042, 1962.

\bibitem{Yang_1}
Pengcheng Yang, Martin~B Plenio, and Jianming Cai,
\newblock ``Dynamical nuclear polarization using multi-colour control of color
  centers in diamond'',
\newblock {\em EPJ Quantum Technology}, vol. 3, pp. 1--9, 2016.

\bibitem{Buks_QMLN}
Eyal Buks,
\newblock {\em Quantum mechanics - Lecture Notes},
\newblock http://buks.net.technion.ac.il/teaching/, 2020.

\bibitem{Bassi_055027}
Angelo Bassi and Kasra Hejazi,
\newblock ``No-faster-than-light-signaling implies linear evolution. a
  re-derivation'',
\newblock {\em European Journal of Physics}, vol. 36, no. 5, pp. 055027, 2015.

\bibitem{Jordan_022101}
Thomas~F. Jordan,
\newblock ``Assumptions that imply quantum dynamics is linear'',
\newblock {\em Phys. Rev. A}, vol. 73, pp. 022101, Feb 2006.

\bibitem{Penrose_4864}
Roger Penrose,
\newblock ``Uncertainty in quantum mechanics: faith or fantasy?'',
\newblock {\em Philosophical Transactions of the Royal Society A: Mathematical,
  Physical and Engineering Sciences}, vol. 369, no. 1956, pp. 4864--4890, 2011.

\bibitem{Leggett_939}
A.~J. Leggett,
\newblock ``Experimental approaches to the quantum measurement paradox'',
\newblock {\em Found. Phys.}, vol. 18, pp. 939--952, 1988.

\bibitem{Leggett_022001}
A.~J. Leggett,
\newblock ``Realism and the physical world'',
\newblock {\em Rep. Prog. Phys.}, vol. 71, pp. 022001, 2008.

\bibitem{Schrodinger_807}
E.~Schrodinger,
\newblock ``Die gegenw¨artige situation in der quantenmechanik'',
\newblock {\em Naturwissenschaften}, vol. 23, pp. 807, 1935.

\end{thebibliography}

\end{document}